\documentclass[onecolumn,sort&compress,numbers]{els-mrw} 

\usepackage{amsmath,amssymb,amsfonts,amsthm,makeidx,graphicx}
\usepackage{txfonts}
\usepackage{helvet}
\usepackage{hyperref}

\begin{document}


\chapter{Exotics with gluonic excitations}\label{chap1}

\author[1]{Jozef Dudek}%
\address[1]{\orgname{The College of William \& Mary}, \orgdiv{Department of Physics}, \orgaddress{Williamsburg, VA 23187, USA}}

\articletag{A pedagogic introduction to the physics of gluonic excitations within the hadron spectrum.}

\maketitle

\begin{abstract}[Abstract]
That the gluons of quantum chromodynamics self-interact with a strong coupling suggests the possibility of hadrons in which the gluonic field plays an essential role: \emph{Glueballs} in which gluons stick together to produce a hadron in which no quarks are required, and \emph{hybrids} in which an excited gluonic field couples to quarks. 
This chapter presents our contemporary theoretical understanding of such states, and the challenges associated with definitively identifying their presence within the experimental spectrum of hadrons.
\end{abstract}

\begin{keywords}
 	Glueballs\sep Hybrids\sep QCD \sep Exotics \sep Spectroscopy
\end{keywords}

\section*{Objectives}
\begin{itemize}
	\item Introduce the possibility that the hadron spectrum contains states that are sensitive to the gluonic field in quantum chromodynamics.
	\item Describe what is currently known about \emph{glueballs}, \emph{hybrid mesons} and \emph{hybrid baryons} in QCD and in experiment.
	\item Describe how we might come to an understanding of these states through experimental and theoretical advances.
\end{itemize}

\section{Introduction}\label{sec:intro}

Systematic patterns in the spectrum of hadrons observed in experiments in the $20^\mathrm{th}$ century led to the \emph{quark model}, in which mesons are described as bound-states of a quark and an antiquark ($q\bar{q}$), while baryons contain three quarks ($qqq$)\footnote{Presented in more detail in Chapter 20008 ``Constituent quark model'' [\url{https://arxiv.org/abs/2504.07897}].}. This ultimately led to the development of quantum chromodynamics (QCD) as the fundamental theory underlying the physics of hadrons, in which color-charged quarks interact with color-charged gluons, which also self-interact. At low energies these interactions are all strong, and not adequately described by perturbation theory, with quarks and gluons empirically being confined within hadrons. The quark model as a characterization scheme has remained a remarkably robust tool for describing the experimental spectrum, an observation which looks peculiar given the properties of QCD, where any color-singlet combination of quarks and gluons is, in principle, allowed. The light ($u,d$) and strange ($s$) quarks have masses that are much lower than many observed hadron masses, so adding additional $q\bar{q}$ pairs would seem to be quite possible on energetic grounds, generating `multiquark' states that might have flavor quantum numbers outside of those allowed in the quark model\footnote{Such states are discussed in Chapter 20025 ``Hadronic molecules and multiquark states'' [\url{https://arxiv.org/abs/2504.06043}].}. Another possibility is that the gluonic field, which is strongly coupled to itself and to quarks, could have a spectrum of excitations that might manifest as additional hadron states lying outside the quark model. It is these hypothetical hadrons, known as \emph{glueballs} if no quarks are involved, and \emph{hybrids} if the gluonic excitation is coupled to quarks, that we will discuss in the current chapter.

\smallskip

The primary challenge in testing the hypothesis of gluonic excitations lies in uniquely identifying an experimentally observed hadron as having glueball or hybrid structure, and indeed, given the strongly-coupled nature of QCD, determining whether physical eigenstates are actually admixtures of multiple different internal structure basis states. In practice most hadrons beyond the lightest few are short-lived \emph{resonances} which decay rapidly into one or more final states made up of several lighter hadrons, and what can be extracted from experimental measurements are flavor and spin-parity quantum numbers, masses and decay widths, branching fractions to various final states, and couplings to particular production processes. A phenomenological approach is to use these observations to build a picture of the possible internal structure of each hadron in the spectrum. An important special case arises from the fact that the assumed $q\bar{q}$ structure of mesons within the quark model does not generate all possible $J^{PC}$ quantum numbers, so observation of a state with $J^{PC}$ outside the allowed set, a so-called \emph{exotic}, would indicate some type of novel structure. 

\smallskip

The phenomenological approach relies upon there being some expectations for how different quark-gluon structures might be produced experimentally, their expected masses, and expected decay properties. In the past these expectations were generated within \emph{models}, typically inspired by the structure of QCD, in which explicit calculations could be carried out, something that was not practical within QCD at the time. 
As well as model expectations, another common approach is to make use of (approximate) symmetry arguments, particularly the  SU(3) flavor symmetry relating $u,d,s$ quarks, and the empirically inferred `OZI-rule' which suggests (within a specific quark model interpretation) that decays in which the meson $q\bar{q}$ annihilate and new $q\bar{q}$ pairs are produced are suppressed relative to decays where the initial $q\bar{q}$ pair is retained.

\smallskip

The (relative) success of the quark model in describing the resonance content of older experimental data coming from a limited set of production processes inspired a `lore' that certain underexplored production processes might be more favorable for production of gluonic hadrons. These would be processes that are, in some sense, `gluon rich', by virtue of needing to go through a gluonic intermediate state -- an example would be charmonium decays to light hadrons in which the $c\bar{c}$ pair must annihilate into gluons. Whether this argument has validity, given the strong coupling between gluons and quarks, is an open question. Many `gluon rich' processes have now been measured, and they have not proven to be the gluonic hadron factories they were once hoped to be, but considered as part of a broad spread of production processes, the data they provided has been very useful as a tool to help constrain the phenomenological approach.

\smallskip

An important item of progress on the theoretical side has been the development of lattice QCD as a tool for first-principles computation in QCD\footnote{See Chapter 20019 ``Lattice QCD calculations of hadron spectroscopy'' [\url{https://arxiv.org/abs/2505.10002}]}. Clear evidence for a detailed spectrum of glueballs has been obtained within calculations of SU(3)--color Yang-Mills theory, \emph{i.e.} QCD without quarks, and inferences made from this spectrum have dominated phenomenological thinking about glueballs for more than 20 years. Calculations within QCD using artificially heavy quarks, crudely ignoring the hadronic decays of excited states, have provided hadron spectra that reproduce many of the systematics present in the experimental spectrum and the $q\bar{q}$, $qqq$ quark model, but also indications for states that appear to have \emph{hybrid} characteristics. In recent years, pioneering lattice QCD calculations of mesons as \emph{resonances} in meson-meson scattering and production processes have appeared, allowing first-principles determination of masses, widths, and branching fractions.

\smallskip
Since glueballs within QCD are expected to have the same quantum numbers as `conventional' $q\bar{q}$ mesons, experimental searches have focussed on identifying `supernumerary' states, \emph{i.e.} finding mass regions in the spectrum of hadrons where there are more resonances observed than expected on the basis of the $q\bar{q}$ quark model. Guidance from the computed Yang-Mills spectrum leads to consideration of scalar ($0^{++}$), pseudoscalar ($0^{-+}$), and tensor ($2^{++}$) quantum numbers, which are expected to be lightest. 
Most attention has landed on the possibility that three scalar resonances, $f_0(1370),\, f_0(1500),\, f_0(1710)$, believed to contribute to experimental data in several different production mechanisms, should contain a glueball component given that only two states would be expected in this energy region in a $q\bar{q}$ model. Schemes in which the three states are each admixtures of $u\bar{u} + d\bar{d}$, $s\bar{s}$, and glueball basis states have been proposed, and the mixing picture constrained by measured decays and production strengths subject to certain model assumptions.

\smallskip
A single candidate state, the $\pi_1(1600)$, a resonance with exotic $J^{PC} = 1^{-+}$ quantum numbers, has dominated experimental studies of hybrid mesons for more than thirty years. Recent analyses of experimental data, particularly data from the COMPASS experiment
, appear to have resolved a longstanding mystery where \emph{two} isovector $1^{-+}$ states seemed to be present, with an additional lighter $\pi_1(1400)$ state being present in just a single decay mode (in place of the $\pi_1(1600)$), with it now seeming likely that the lattice QCD prediction of a single such state is reflected in experiment. Very recently, a candidate to be an isospin partner of this state, the $\eta_1(1855)$, has been claimed in charmonium radiative decays at BESIII
. There are no convincing candidates for hybrid \emph{baryons} to date, with the fact that all half-integral $J^P$ values can be reached with quark model $qqq$ constructions meaning that there are no exotic quantum numbers to explore.

\smallskip
In this chapter we will consider, in turn, glueballs, hybrid mesons, and hybrid baryons, discussing what is known about each within QCD (primarily using the results of lattice QCD computations) and what has been inferred from experimental measurements. We will conclude with a brief summary and some projections for likely progress in the near future.

\newpage
\section{Glueballs}\label{sec:glueballs}

\begin{figure}[h]
	\centering
	\includegraphics[width=.5\textwidth]{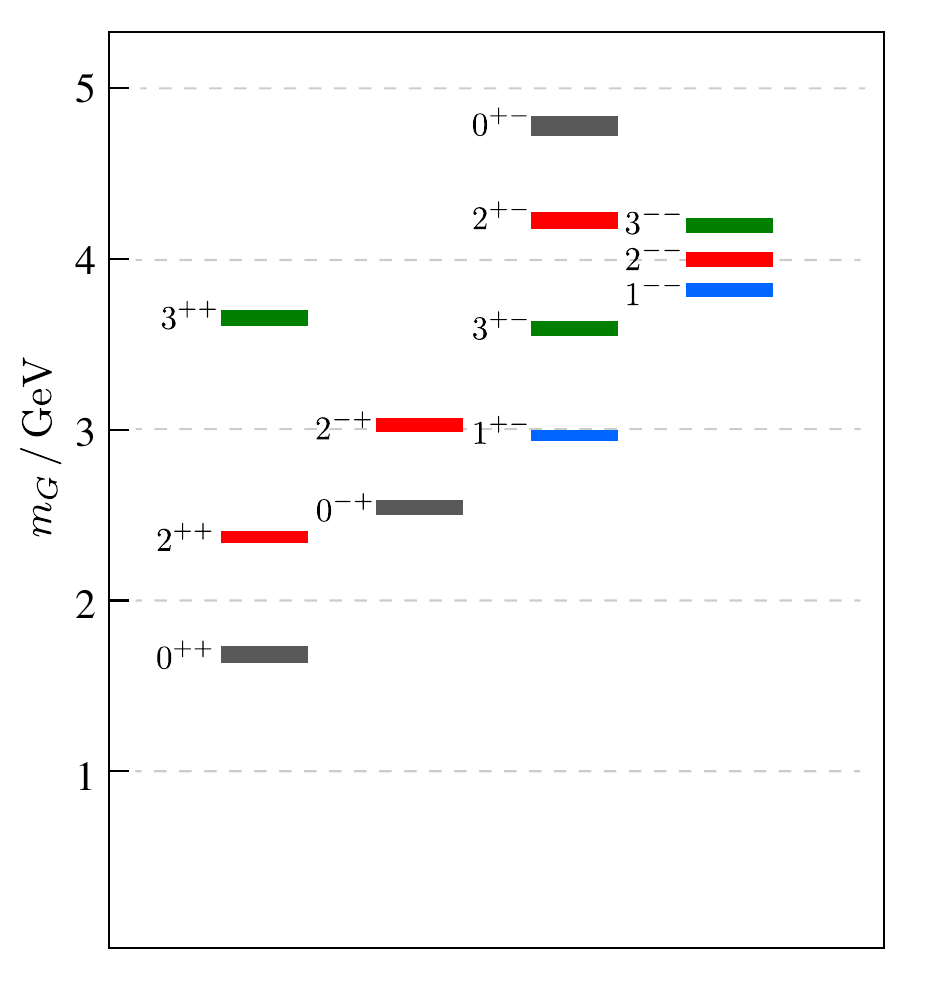}
	\caption{Spectrum of color-singlet glueballs in a lattice calculation of SU(3) Yang-Mills, quarkless-QCD, labelled by $J^{PC}$ quantum numbers. Figure adapted from Ref.~\cite{Chen:2005mg}, which provides an update of the earlier pioneering calculation of Ref.~\cite{Morningstar:1999rf}.	
	}
	\label{fig:glueballs}
\end{figure}

The strong self-coupling of the gluonic field in QCD pushes us towards the possibility of a spectrum of states of bound glue without quarks needing to be involved. In fact, a cleaner place to study such states is in the theory of QCD without quarks, known as SU(3) Yang-Mills or sometimes `gluodynamics'. This theory can be studied non-pertubatively in numerical calculations by placing it on a space-time grid, and evaluating correlation functions as an average over Monte-Carlo sampled gluon field configurations. An example of the results of such a calculation, presenting the spectrum of color-singlet states extrapolated to the limit where the lattice grid spacing goes to zero, is given in Figure~\ref{fig:glueballs}. We observe a detailed spectrum of states, with the lightest few having scalar ($0^{++}$), tensor ($2^{++}$), and pseudoscalar ($0^{-+}$) quantum numbers. 
Within this quarkless theory, these glueballs are the only mesons, and as such all states with a mass below around 3.5 GeV are stable, while some higher-lying states might decay into a pair of lighter glueballs.

These numerical results indicate that the concept of a spectrum of bound-states of gluons within strongly-coupled SU(3) gauge theory is solid, but the question then becomes how these observations inform our expectations of how glueballs might manifest \emph{in QCD} when quark degrees of freedom are also present, and strongly coupled to the gluonic field?

\smallskip
True glueballs are obviously electrically uncharged and carry none of the quantum numbers associated with quark flavor. As such they are isospin=0 states, and for example the pseudoscalar glueball has the same flavor and $J^{PC}$ quantum numbers as the well-studied $\eta$, $\eta'$ mesons, or any of several observed heavier resonances with $I=0$, $J^{PC}=0^{-+}$.
This observation leads to the principal challenge in identifying the role of glueballs in the hadron spectrum: ``how do we tell them apart from `ordinary' mesons?'' and indeed, pushing the concept further, what even do we mean by `ordinary meson' if glueball components can be present?

\smallskip
Historically the community has been guided by the empirical success of the quark model, which proposes a $q\bar{q}$ structure for mesons, predicting a spectrum of states lacking certain $J^{PC}$ possibilities, and a limited set of flavor quantum numbers. Adding in some dynamical assumptions, a mass ordering of $J^{PC}$ in the spectrum can be proposed that is in reasonable agreement with experiment. Further details are presented in Chapter 20008 ``Constituent quark model'' [\url{https://arxiv.org/abs/2504.07897}].
One possible realization would be that a glueball spectrum somewhat like the Yang-Mills spectrum appears, simply \emph{augmenting} the spectrum of $q\bar{q}$--like mesons. In this case there would be `excess' states relative to the counting predicted by the quark model. Since the glueball states would lie above thresholds to decay into systems of light mesons, unless some quirk of QCD prevents them decaying, they would appear as resonances, just as heavier $q\bar{q}$--like states would. One might hope to identify them as glueballs by some unique feature, or combination of features, of their decay or production characteristics, but only if some dynamical understanding of these processes could be reached.

\smallskip
Another, perhaps more likely, possibility is that the true spectrum is that of admixed states which are not purely glueball or $q\bar{q}$--like, but rather superpositions of these viewed as basis states. Quarks and gluons are strongly coupled within QCD, so there seems to be no obvious reason why a glueball basis state would not mix strongly with an isoscalar $q\bar{q}$ basis state through the non-perturbative dynamics of QCD. In this case, we still expect more states than the $q\bar{q}$ picture would suggest, but any characteristic property of a glueball (if any such thing exists) would be distributed across all the physical eigenstates that contain a glueball component. 

\smallskip
Unfortunately, despite much effort, we lack significant first-principles understanding of how hadronic decay proceeds in QCD, such that most analyses resort to approximate symmetry arguments, or assumptions about dynamics within models.  
Most general symmetry arguments for glueball nature (\emph{e.g.} Ref.\cite{Lipkin:1981ak}) are really establishing that the state transforms as an SU(3) flavor singlet, which a pure glueball of course would be (as it knows nothing of quark flavor directly), but it is perfectly possible to construct a $q\bar{q}$ state that is an SU(3) flavor singlet, $u\bar{u} + d\bar{d} + s \bar{s}$, indeed the $\eta'$ meson is believed to have a flavor structure very close to this.

\smallskip
The empirical OZI--rule commonly forms part of the phenomenology, where decays that require quark annihilation are suppressed relative to those in which only new $q\bar{q}$ pairs are added (all within a  simple $q\bar{q}$ quark-model interpretation of the hadrons involved). An important example is the decay of the $\phi(1020)$ meson, where $\phi \to \pi \pi \pi$ is significantly suppressed relative to $\phi \to K\overline{K}$, despite having a much larger phase-space. The conventional interpretation is that the $\phi$ has a dominant $s\bar{s}$ structure so that the $\pi\pi\pi$ decay requires annihilation of the strange quarks, while the $K\overline{K}$ decay does not. Another empirical example is the assignment of $u\bar{u}+d\bar{d}$ structure to the tensor meson $f_2(1270)$ which has dominant $\pi\pi$ decays, and $s\bar{s}$ structure to $f_2(1525)$ which mainly decays to $K\overline{K}$~\footnote{A similar pattern is observed in lattice QCD calculations of these states (at heavier than physical light quark masses) in Ref.~\cite{Briceno:2017qmb}.}. 
A consequence of the OZI--rule is the commonly made assumption that isoscalar $q\bar{q}$ mesons are always `ideally' flavor mixed, as $u\bar{u}+d\bar{d}$, and $s\bar{s}$, without significant hidden-flavor mixing.
Despite being an often--used tool in interpretation of experimental meson spectra, the dynamical origin of the OZI--rule is poorly understood, and it does not seem to apply in all $J^{PC}$ channels, notably the pseudoscalar $\eta$ and $\eta'$ which do not appear to be ideally flavor mixed, rather being closer to SU(3) octet and singlet respectively, and possibly also the axial $f_1$ mesons, for which there is some evidence of modest flavor mixing (\emph{e.g} in Ref.~\cite{Close:1997nm}).

\smallskip
As well as interpretational challenges, the search for glueballs also has to face the difficulty of uniquely and reliably identifying from experimental data a spectrum of overlapping resonances which may be broad. Experimentally we identify the presence of resonances and their quantum numbers by examining the invariant mass and angular distributions of their decay products, and while isolated sharp peaks in invariant mass are clear signals for narrow resonances, broad overlapping states can manifest as in quite complicated ways that simple analyses may misidentify. As we will see below, in the scalar, tensor and pseudoscalar sectors, a universally accepted spectrum of resonances below \mbox{2.5 GeV} has not yet emerged.

\bigskip
Chapter 20002 ``Key Historical Experiments in Hadron Physics'' [\url{https://arxiv.org/abs/2503.14689}] presents a summary of some of the main experiments that have influenced our understanding of the hadron spectrum. Much of the phenomenology that attempts to infer the presence of glueballs in the meson spectrum is grounded in analysis of data from $p\bar{p}$ annihilation at LEAR, with significant contemporary input coming from BESIII, mainly in $J/\psi$ decays to light mesons, especially radiative decays. Charmonium decays are commonly thought of as being `glue-rich' since the $c\bar{c}$ must annihilate to an intermediate gluon system in order to produce light mesons.
In principle two--photon couplings of meson resonances, typically extracted from $\gamma \gamma$ \emph{production} (in $e^+ e^-$ machines), serve as a filter against pure--glueball nature owing to the lack of electrically charged constituents in a pure glueball. But in a strongly-coupled system this logic is not so clear: if the glueball is able to decay to an electrically charged meson pair, we generically expect the two-photon initial state to be able to couple to that meson pair (via a Born-term), and then to rescatter through the glueball resonance\footnote{A comprehensive analysis of high-statistics $\gamma \gamma$ data from BESIII for coupled final states, $\pi\pi$, $K\bar{K}$, $\pi \eta$ is nearing completion~\cite{Kussner:2024kmh}.}.

\begin{figure}[h]
	\centering
	\includegraphics[width=.9\textwidth]{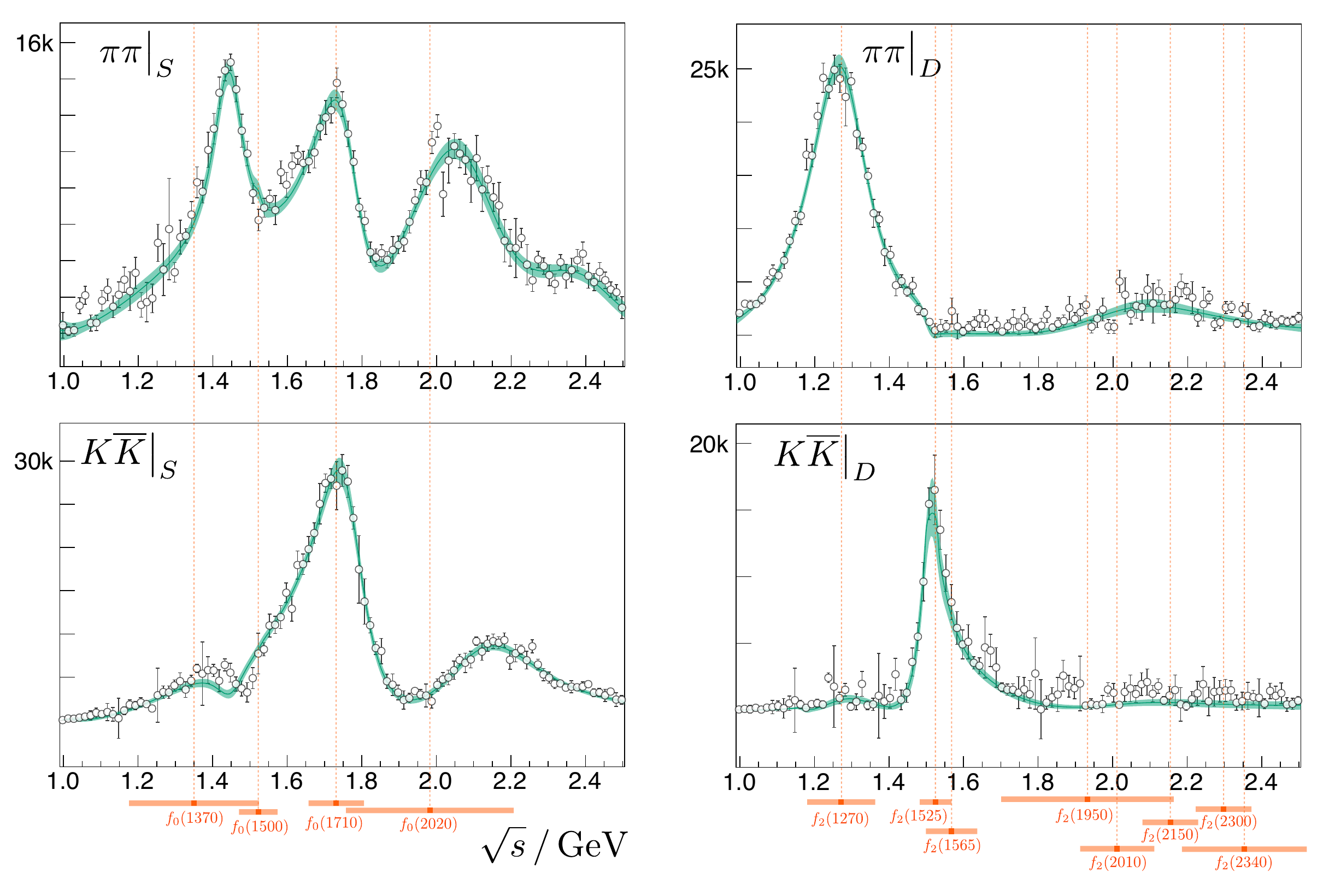}
	\caption{ BESIII data~\cite{BESIII:2018ubj, BESIII:2015rug} on $J/\psi \to \gamma \, \pi\pi$ and 	$J/\psi \to \gamma \, K\overline{K}$, number of events as a function of $\pi \pi$ or $K\overline{K}$ invariant mass. Left panels have final-state meson system with $J^{PC}=0^{++}$, right panels with $J^{PC} = 2^{++}$. Orange lines indicate the mass of each established isoscalar scalar and tensor meson listed in the PDG Review of Particle Physics~\cite{ParticleDataGroup:2024cfk}, with the total width of each resonance indicated by the orange band below. 
	Coupled-channel description of Ref.~\cite{Rodas:2021tyb} shown by green curves -- this fit finds no pole corresponding to $f_0(1370)$, narrow poles that agree reasonably with the $f_0(1500)$ and $f_0(1710)$, and broader poles $f_0(2020)$, $f_0(2330)$. In the tensor sector, only $f_2(1270), f_2(1525)$ and a broad $f_2(1950)$ feature.
		}
	\label{fig:JpsiRadiative}
\end{figure}

\smallskip
The Yang-Mills spectrum presented in Figure~\ref{fig:glueballs} suggests that a scalar ($0^{++}$) glueball could be lightest in QCD, and most phenomenological attention has been directed to this sector. Understanding states with these quantum numbers is not straightforward, with even the lightest two resonances, the $\sigma/f_0(500)$ and the $f_0(980)$ being the subject of a great deal of discussion over many years regarding their structure, see Chapter 20021 ``Light meson resonances'' [\url{https://arxiv.org/abs/2509.08648}]. Generally it is assumed that the impact of glueball basis states will be within scalar resonances heavier than these.
Figure~\ref{fig:JpsiRadiative}, showing recent BESIII data on the charmonium radiative decays $J/\psi \to \gamma \, (\pi\pi, K\overline{K})$, illustrates the challenge of pinning down the excited isoscalar scalar and tensor meson resonance spectrum. The orange lines indicate the locations of `established' resonances listed in the PDG~\cite{ParticleDataGroup:2024cfk} lying between 1 GeV and 2.5 GeV, and we see that except in limited cases ($f_0(1710) \to \pi \pi, K\overline{K}$, $f_2(1270) \to \pi\pi$, $f_2(1525) \to K \overline{K}$), the high-statistics radiative decay data does not peak at the supposedly established resonance locations. The data does have \emph{other} peaks, but these do not line up with the `established' resonances, which is perfectly possible for broad overlapping states, indicating the need for careful analysis. The curves shown in the figure correspond to a coupled-channel fit~\cite{Rodas:2021tyb} to the data, apparently successfully describing it in terms of fewer resonances than the number listed in the PDG (particularly in the tensor spectrum). On the other hand, when Klempt \emph{et al}~\cite{Sarantsev:2021ein} fit these data together with phase-shift data and $\bar{p}N$ data they claim \emph{eight} scalar resonances\footnote{Kopf \emph{et al}~\cite{Kopf:2020yoa} use a similar method and data set to Klempt \emph{et al} (excluding the radiative decay data) but do not seem to require as many scalar resonances.} between 1 GeV and 2.5 GeV.
Clearly, determining the resonance content of the isoscalar scalar and tensor sector is complicated and requires careful and highly constrained analysis, and the community has not yet converged to a consensus.

\smallskip
Nevertheless, despite these caveats, 
it is commonly accepted that for the scalars above 1 GeV and below 2 GeV, there are probably three resonances, a broad $f_0(1370)$ lying below narrower $f_0(1500)$, $f_0(1710)$ states. In the most prevalent interpretation this is one more state than would be expected in the $q\bar{q}$ quark model. Various admixture schemes have been proposed using $u\bar{u} + d\bar{d}$, $s\bar{s}$, and glueball as the basis states of which the three resonances are superpositions. In one phenomenological interpretation of experimental data~\cite{Amsler:2018zkm, Close:2001ga}, the $f_0(1370)$ is dominated by $u\bar{u} + d\bar{d}$, the $f_0(1710)$ is dominated by $s\bar{s}$, and it is the $f_0(1500)$ which contains the largest glueball component. This interpretation is not unique; different mixing patterns for these three states are discussed in the review article, Ref.~\cite{Crede:2008vw}. There are also alternative pictures which consider more than these three states, including at least one in which it is heavier states that have dominant glueball content~\cite{Sarantsev:2021ein}.

\smallskip
In the tensor sector, the lighest two states, the $f_2(1270)$ and $f_2(1525)$ are well established resonances, and have properties in excellent agreement with expectations for ideally--flavor--mixed $q\bar{q}$ states. The spectrum of resonances above these two is not sufficiently well established to realistically offer any reliable interpretation in terms of glueball content.

\smallskip
A pseudoscalar glueball search might be expected to be quite challenging owing to how heavy the state is anticipated to be. The lightest $0^{-+}$ glueball in the quarkless Yang-Mills spectrum, shown in Figure~\ref{fig:glueballs}, lies near 2.5 GeV, and if we assume a similar mass in quarkfull QCD, such a state would be heavier than many other anticipated resonances. As an example, Figure~\ref{fig:LQCDmesons} shows a lattice QCD spectrum with heavier than physical $u,d$ quarks in which below 2.5 GeV there are \emph{six} isoscalar pseudoscalar states. As well as the $\eta$ and $\eta'$, there are a pair of states which might be interpreted as $q\bar{q}$ radial excitations, and then a pair of states which appear to be \emph{hybrid mesons}. Upon lowering the light quark mass, even more non-glueball states could be lighter than the glueball, and all this discussion is before we consider mixture of basis states, and the possibly large decay widths\footnote{Which are not captured in the calculations leading to Figures~\ref{fig:glueballs} and \ref{fig:LQCDmesons}.} that these states could have.

\smallskip
Setting aside some historical confusion\footnote{The ``$\iota(1450)$'' story is reviewed in Ref.\cite{Godfrey:1998pd}.} about the experimental isoscalar $0^{-+}$ resonances above 1 GeV, the excited spectrum is still not well established, and neither is there any simple phenomenological picture than is generally agreed upon. There has been a recent proposal~\cite{BESIII:2023wfi} that a resonance, $\eta(2370)$, produced in $J/\psi$ radiative decays, and seen decaying to $K_S K_S \eta'$, is a strong candidate to be the pseudoscalar glueball (it was also probably seen earlier decaying to $\pi \pi \eta'$). This claim seems to be largely based upon its mass and perhaps some branching fraction limits inferred from lattice QCD that will be discussed below. At the current time, identifying this state as a glueball would seem to be rather speculative.

\bigskip

Theoretical progress is required if we are to come to definitive conclusions about the role of glueballs in the meson spectrum. A high priority is to develop and apply sophisticated coupled-channel amplitude analysis constrained by as much accumulated experimental data as possible in order to establish the resonance content of several $J^{PC}$ quantum numbers. The hope is that such analyses will converge to a unique spectrum of states including information about their branching fractions and production couplings. Having reliable spectral information is a precondition to attempting any kind of phenomenological investigation that might expose the role of glueballs.

\smallskip
Further developments are also required to get a faithful picture of the expected meson spectrum in QCD. Some recent progress includes lattice QCD computations of a spectrum of states analogous to Figure~\ref{fig:glueballs}, but on a background of gauge field configurations computed with quarks `in the sea', rather than in Yang-Mills (see Ref.~\cite{Vadacchino:2023vnc} for  summary). It is not clear how one should interpret these results, given that lattice QCD should yield a fairly dense spectrum which depends upon the finite spatial volume of the lattice, one which can be related to scattering amplitudes and resonances having hadronic decays~\cite{Briceno:2017max}. This spectrum should include \emph{all} mesons, not just glueballs -- further discussion appears in the subchapter on hybrid mesons which follows this one.
Other recent lattice work estimates rates for charmonium radiative decay to glueballs, $J/\psi \to \gamma\,  G$, in Yang-Mills (no light quarks, no glueball decays) with charm-quarks added on the gluon background (see Ref.~\cite{Gui:2019dtm} and references therein). These results have been used in interpretations of BESIII radiative decay data, but since it is not clear how (or even if) the glueballs in QCD with quarks are related to the Yang-Mills glueballs, how rigorous these interpretations are is up for discussion.

\smallskip
A new $p\bar{p}$ experiment, using a modern spectrometer known as PANDA, has been proposed to be part of the FAIR project in Darmstadt, Germany. As part of its broad set of aims, this experiment will be capable of providing new high-quality meson production data that is expected to have a significant influence on the phenomenological approach to glueball searches.

\smallskip
Many decades after their proposal, glueballs continue to inspire significant experimental and theoretical attention, but a clear understanding thus far eludes us. We will return to what can be done in the future after we discuss hybrid mesons and baryons.

\section{Hybrid Mesons}\label{sec:hybridmesons}

While glueballs, as states of pure glue, are emblematic of the gauge-field self-coupling in QCD, they are not the only possible novel states containing an excitation of the gluonic field, and indeed, as described above, they may not be the easiest to determine experimentally. Within QCD, the gluonic field is as strongly coupled to quarks as it is to itself, and this raises the possibility of hadrons containing quarks alongside the gluonic field in an essential structural role. These are usually called \emph{hybrid} hadrons, and in the meson case, they potentially have properties that would make them easier to identify in experiment.

\smallskip
An obvious way that a \emph{hybrid meson} could differ from a `conventional' meson in the simple $q\bar{q}$ quark model is if the $q\bar{q}$ pair is in a color \emph{octet}, rather than a color singlet, with this color neutralized overall in the meson by the presence of a color octet gluonic excitation.
The additional quantum numbers introduced by the gluonic excitation potentially allows for the presence of \emph{exotic} $J^{PC}$ quantum numbers for the meson such as $0^{--}, 0^{+-}, 1^{-+}, 2^{+-} \!\ldots\,$, which are exotic in the sense that they cannot appear for a $q\bar{q}$ bound state. $J^{PC}$ quantum numbers of resonances can be established experimentally by examining angular dependences of the multiparticle final states when the excited hadron decays, and if a resonance is observed with exotic $J^{PC}$, it must go beyond the simple quark model, and a hybrid nature is a plausible hypothesis. 

\smallskip
With techniques for first-principles calculation in QCD being limited historically, early predictions came from models motivated by QCD in which the gluonic excitation was assumed to take one of several plausible forms. One possibility was that, in analogy to the large effective mass of quarks in the `constituent' quark model (see Chapter 20008 ``Constituent quark model'' [\url{https://arxiv.org/abs/2504.07897}]), the gluon could be assumed to obtain an effective mass through self-interactions, and behave as a heavy `constituent gluon'. Presence of a constituent gluon alongside quarks would generate a hybrid hadron. In another picture, confining the gluonic field within a spherical `bag' led to possible excited field modes with an associated energy that could appear alongside quarks in a hybrid hadron. Treating a spatially separated quark--antiquark pair as being connected by a narrow tube of gluonic flux allowed for transverse oscillatory excitations of that tube, and an associated set of hybrid mesons. A discussion of these models, somewhat contemporary to their development, can be found in Ref.~\cite{Close:1987er}, and a more modern consideration appears in Ref.~\cite{Meyer:2015eta}. These models gave differing predictions for the expected spectrum of hybrid mesons, both in the set of expected $J^{PC}$ states, and in the anticipated mass spectrum. As well as exotic $J^{PC}$ states, they typically also predicted states with quantum numbers also accessible to a simple $q\bar{q}$ pair.

\smallskip
Should an exotic $J^{PC}$ resonance be seen in experiment, ruling out that the state has a structure of a larger number of quark constituents, such as $qq\bar{q}\bar{q}$, requires further examination. In general such multiquark pictures lead to at least some number of combinations of quarks which have exotic \emph{flavor} quantum numbers not accessible to $q\bar{q}$ (such as isospin $> 1$ or magnitude of strangeness $>1$), suggesting that we should examine patterns of states with different flavor. Conversely, hybrid mesons would appear in the same flavor octets and singlets as conventional $q\bar{q}$ mesons, as the gluonic excitation does not introduce any additional flavor.

\smallskip

To date the largest experimental focus has been on the exotic $1^{-+}$ possibility. An isospin $I=1$ state with these quantum numbers would be denoted $\pi_1$ in the modern PDG naming scheme, and could potentially appear through decay to $\pi \eta$ or $\pi \eta'$ in a $P$--wave, a relatively simple final state. Presence of a $1^{-+}$ state somewhere in the predicted spectrum was a typical feature of the models outlined above.

\medskip

More recently, developments in lattice QCD have made it possible to theoretically explore the meson spectrum more directly within QCD. As described in Chapter 20019 ``Lattice QCD calculations of hadron spectroscopy'' [\url{https://arxiv.org/abs/2505.10002}], correlation functions can be computed using a basis of operators with the quantum numbers of mesons, constructed out of quark and gluon fields, and a matrix formed. Diagonalization of the matrix gives access to the mass spectrum of excited states, along with a numerical measure of how well each operator overlaps with each state.

\begin{figure}[h]
	\centering
	\includegraphics[width=.95\textwidth]{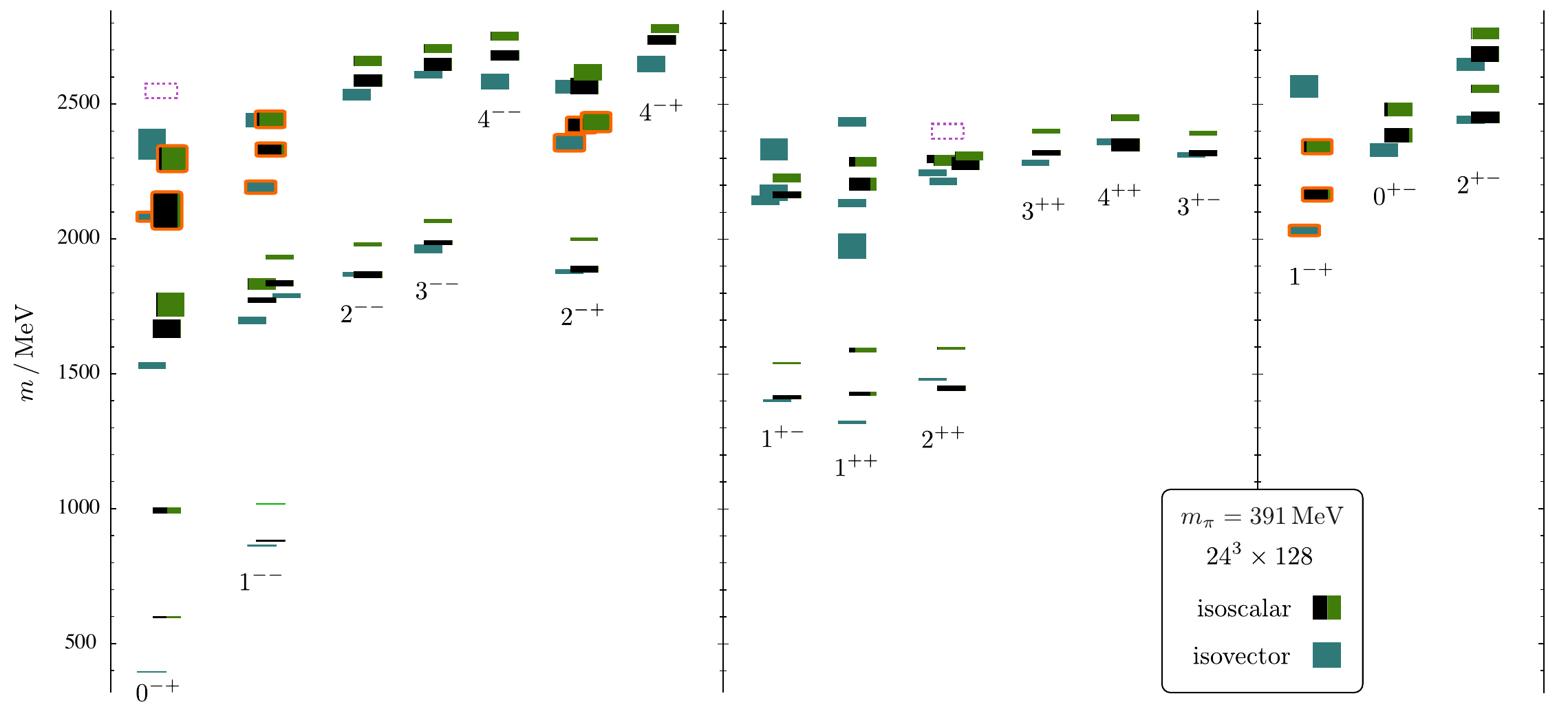}
	\caption{Spectrum of light meson states computed in lattice QCD with light quarks such that $m_\pi = 391$ MeV. Isoscalar states ($I=0$) have their hidden-light ($u\bar{u}+d\bar{d}$) and hidden strange ($s\bar{s}$) content shown by the relative sizes of black/green in the box. States found to have large overlap with operators featuring a chromomagnetic structure are highlighted with an orange border, which serve as the proposed lightest hybrid meson supermultiplet. No glueball-like operators were included in the isoscalar basis used in this calculation, but for illustration the locations of the lightest $0^{-+}$ and $2^{++}$ states found in the quarkless calculation of Ref.~\cite{Chen:2005mg} are shown by the open purple boxes. 
Figure adapted from Ref.~\cite{Dudek:2013yja}.	
	}
	\label{fig:LQCDmesons}
\end{figure}

\smallskip
Figure~\ref{fig:LQCDmesons} shows the result of such a calculation~\cite{Dudek:2013yja}, performed with a version of QCD in which strange quarks are assigned approximately their physical mass, while the degenerate $u,d$ quarks are chosen to  have a mass larger than their physical value. In this case the pion is found to have a mass of 391 MeV.
An operator basis of fermion bilinears, $\bar{\psi} \Gamma \overleftrightarrow{D} \!\ldots\! \overleftrightarrow{D} \psi$, was used, with up to three gauge-covariant derivatives, leading to a large number of operators with each $J^{PC}$ shown in the figure. 
A separate calculation was done for the $I=1$ spectrum, where $q\bar{q}$ annihilation is not possible, and the $I=0$ spectrum where it is.  In the second case the operator basis contains both hidden-light quark ($u\bar{u} + d\bar{d}$) and hidden-strange quark ($s\bar{s}$) operators -- the dynamics of QCD in the calculation will decide if these mix significantly in opposition to the naive OZI--rule.

\smallskip
The spectrum shown in the figure indicates that many of the features of the experimental meson spectrum are reproduced in this calculation. The ordering of the lightest states across $J^{PC}$ is in reasonable agreement  with well-established experimental states, and hidden flavor mixing in the isoscalar sector (determined by relative overlap of eigenstates onto $u\bar{u} + d\bar{d}$ versus $s\bar{s}$ operators) aligns with that inferred from experiment. In particular OZI--rule--like ideal flavor mixing (states being close to pure $u\bar{u} + d\bar{d}$ or pure $s\bar{s}$ with little mixing) is observed in most $J^{PC}$, with $1^{--}$ being a relevant example, while strong flavor mixing is observed in $0^{-+}$. Excitations above the ground state are determined in most $J^{PC}$, with some regions of the spectrum having a rather dense set of states (\emph{e.g.} the isoscalar $1^{--}$ sector between \mbox{1.7 GeV} and \mbox{2.0 GeV} where four states appear).

\smallskip
A clear spectrum of \emph{exotic} $J^{PC}$ states can be seen in the rightmost columns, with the lightest such state being an isovector $1^{-+}$, together with its isosinglet partners, lying below heavier positive--parity exotics.
There are also states with non-exotic quantum numbers which appear to lie outside a pattern of states broadly explicable in a $q\bar{q}$ quark model picture. These are highlighted in Figure~\ref{fig:LQCDmesons} with an orange outline, and singling them out for special attention comes from examination of the overlap of these states onto a particular set of operators. The basis includes a number of operators containing the structure
$[\overleftrightarrow{D}_{\!i}, \overleftrightarrow{D}_{\!j}]$, which would be zero for ordinary derivatives, but which in QCD is proportional to the spatial part of the gluonic field strength tensor. The highlighted states have unusually large overlap onto these operators which we interpret as describing $q\bar{q}$ in a color octet coupled to the color octet \emph{chromomagnetic} field.
This examination of overlaps allows us to identify highlighted states with $J^{PC}=$ $0^{-+}$, $1^{--}$, $2^{-+}$ which have masses quite similar to the lightest $1^{-+}$ exotic (which also overlaps strongly onto an operator featuring the chromomagnetic structure). It is argued that these are candidates to make up the lightest hybrid meson \emph{supermultiplet}~\cite{Dudek:2011bn}. This same set of states was found to be present in calculations with a range of light quark masses with pion masses varying from 391 MeV up to 702 MeV, where the $u,d$ quark masses are set to be the same as the physical strange quark mass. The mass gap between the lightest vector meson and the hybrid mesons was found to be largely quark mass independent at around 1.2 GeV (see also Figure~\ref{fig:chromomagnetic}).

\smallskip

A simple picture describing this set of states has the $(q\bar{q})_\mathbf{8}$ system in an $S$--wave, coupled to a $J^{PC} = 1^{+-}$ gluonic excitation of effective mass $\sim 1.2$ GeV, where the total quark spin $S_{\!q\bar{q}}=0$ configuration generates the $1^{--}$ meson, while the $S_{\!q\bar{q}}=1$ configuration gives rise to the $(0,1,2)^{-+}$ states. The quantum numbers inferred for the gluonic excitation are not those of a naive `constituent gluon', and do not match on to predictions from the `flux-tube' model, but do correspond to the lightest confined mode in the `bag' model (further discussion can be found in Ref.~\cite{Dudek:2011bn}).

\begin{figure}[h]
	\centering
	\includegraphics[width=.7\textwidth]{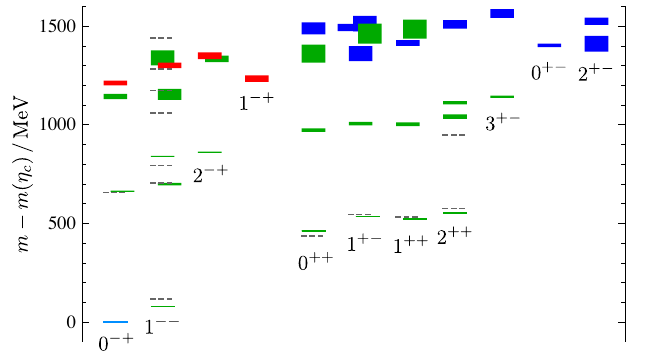}
	\caption{ Spectrum of charmonium mesons computed in lattice QCD with light quarks such that $m_\pi = 391$ MeV. Dashed gray lines show masses of experimental charmonium states listed in the PDG (in 2012). Boxes indicate states extracted from the lattice QCD calculation with the vertical size of the box representing the statistical uncertainty. Green boxes correspond to states interpreted as having conventional $c\bar{c}$ structure while red and blue boxes are identified by their large overlap onto operators containing a chromomagnetic construction. Figure adapted from Ref.~\cite{HadronSpectrum:2012gic}.}
	\label{fig:LQCDccbar}
\end{figure}

\smallskip
The same lattice QCD technology was applied in a calculation with charm-mass quarks, and the more statistically precise signals allowed an extraction of even higher-lying states, as shown in Figure~\ref{fig:LQCDccbar}. As well as the $(0,1,2)^{-+}, 1^{--}$ supermultiplet discussed above (red boxes), the positive parity exotics are found to be partnered by a set of positive parity non-exotics having common operator overlaps (blue boxes). The particular set of $J^{PC}$ observed fits into the chromomagnetic gluonic excitation picture if the $(q\bar{q})_\mathbf{8}$ pair is excited in a $P$-wave. Evidence for another gluonic excitation beyond the lightest chromomagnetic case within these lattice calculations is limited, with an exotic $J^{PC}=0^{--}$ state lying high in the spectrum possibly being an indication of the next excitation.

\smallskip

%
For heavy-quark bound states, the large mass of the quark motivates a different approach, where \emph{potentials} are computed in lattice QCD by measuring the energy of separated static color sources. These are then used to model heavy quark $Q\bar{Q}$ systems in an adiabatic scheme solving a Schr\"odinger equation where the quarks move relatively slowly in the static potential. $Q\bar{Q}$ pairs moving in the lowest potential have a spectrum that matches quark model expectations, but there are also excited potentials corresponding to gluonic excitation, and the spectrum of $Q\bar{Q}$ states in those potentials can be considered to be hybrids (see Ref.~\cite{Braaten:2014qka} and references therein).

\smallskip
It can be argued that the picture described above, giving the lightest supermultiplet extracted from lattice QCD calculations, represents the closest we currently have a reliable prediction of the spectrum of hybrid mesons within QCD, but results such as those in Figure~\ref{fig:LQCDmesons} should be viewed with some caution. Most of these states should actually be short-lived resonances decaying into one or more multi-hadron final state, and as discussed in Chapter 20019 ``Lattice QCD calculations of hadron spectroscopy'' [\url{https://arxiv.org/abs/2505.10002}], resonances appear in the finite volume of the lattice not simply as single energy levels. A plausible interpretation of the spectrum presented in Figure~\ref{fig:LQCDmesons} is that it indicates approximations to the masses of relatively narrow resonances\footnote{See the discussion in Appendix A of Ref.~\cite{Wilson:2015dqa}.}, giving no information about their decays, and the presented spectrum may miss broad resonances.

\bigskip


As mentioned above, the strongest experimental evidence for hybrid mesons is in the $1^{-+}$ isovector channel, dating back to 1988 when a claim was first made for a $\pi_1$ state in the GAMS experiment at the CERN SPS~\cite{IHEP-Brussels-LosAlamos-AnnecyLAPP:1988iqi}. A relatively narrow resonance with a mass near 1400 MeV was proposed to explain $\pi^- p \to (\pi^0 \eta)\, n$ data alongside the dominant $a_2(1320)$ meson. In the four decades since, numerous other experiments examined data containing $\pi \eta$, $\pi \eta'$, and higher multiplicity final states, observing enhancements whose angular dependence is compatible with being due to $1^{-+}$ quantum numbers (a summary as of 2015 can be found in Ref.~\cite{Meyer:2015eta}).

\smallskip
Until recently these data sets appeared to indicate a resonance content that was hard to reconcile with the theoretical picture of hybrid mesons described above. The situation can be illustrated with modern high-statistics data from the COMPASS experiment using a high-energy pion beam which show broad bump structures in both $\pi \eta$ and $\pi \eta'$, but at different locations~\cite{COMPASS:2014vkj}. 
Single-channel Breit-Wigner descriptions of the $\pi \eta$ data suggest a resonance similar to that seen by GAMS, a light but broad $\pi_1(1400)$, while the $\pi \eta'$ data peaks at a higher energy and suggests a $\pi_1(1600)$. Other final states explored in COMPASS show peaking structure compatible with the $\pi_1(1600)$. A similar picture had been seen in earlier experiments, which (until recently) had led to the Particle Data Group listing two exotic states, $\pi_1(1400)$ and $\pi_1(1600)$. 

\smallskip
From the theoretical side, this was a mystery -- lattice QCD studies didn't anticipate two low-lying states, and neither did any of the older plausible models for hybrid mesons. There were phenomenological attempts to accommodate two light resonances by proposing a tetraquark multiplet. (\emph{e.g.} Ref.~\cite{Chung:2003qy}) 
but none of the predicted flavor exotic states have been found in experiment to complete the multiplet\footnote{Meson-meson molecule models can also accommodate two light $1^{-+}$ isovector states~\cite{Yan:2023vbh}.}.

\begin{figure}
	\centering
	\includegraphics[width=.9\textwidth]{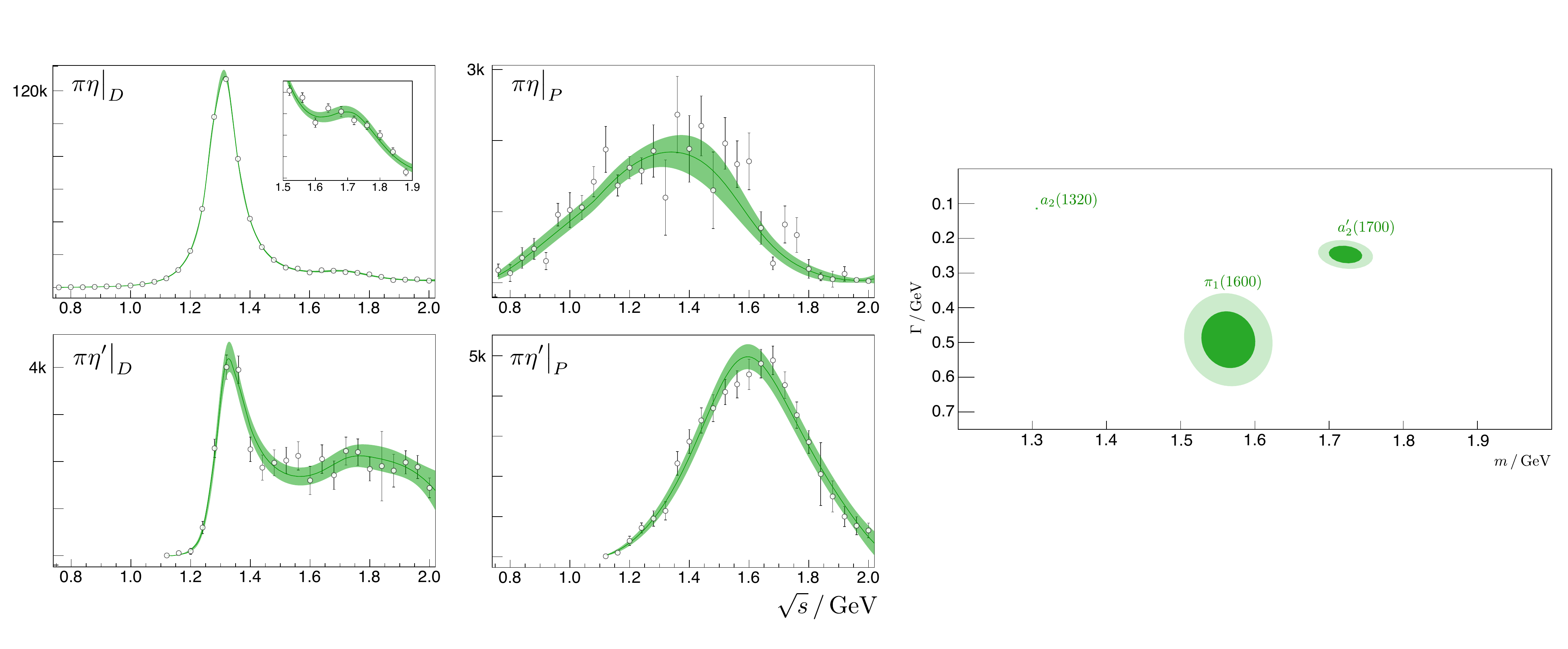}
	\caption{COMPASS $\pi \eta$ and $\pi \eta'$ data described in coupled-channel analysis. $D$-wave ($2^{++}$) amplitudes feature two resonance poles, the narrow, well established, $a_2(1320)$, and a broader $a_2'(1700)$. $P$-wave ($1^{-+}$) amplitudes can be described in terms of a single broad $\pi_1(1600)$ resonance pole.
	 Figure adapted from plots in Ref.~\cite{JPAC:2018zyd}.}
	\label{fig:COMPASS}
\end{figure}

\smallskip
A resolution seems to have been reached recently, where within \emph{coupled-channel} analysis, it has been shown that it is possible to describe both the peaks in $\pi \eta$ \emph{and} $\pi \eta'$ as being due to a single resonance. Ref.~\cite{JPAC:2018zyd} found that the COMPASS data on these two final states could be described by a $\pi_1$ resonance with mass of $1564(24)(86)$ MeV and a rather large width of $492(54)(102)$ MeV, see Figure~\ref{fig:COMPASS}. A somewhat different analysis scheme explored in Ref.~\cite{Kopf:2020yoa} simultaneously described the COMPASS $\pi \eta,\,  \pi \eta'$ data and older Crystal Barrel $p\bar{p}$ data for several final states, as well as $\pi\pi$ phase-shift data -- they also found need for only a single $\pi_1$ resonance with a mass and width statistically compatible with that quoted above.

\smallskip
Older high-energy pion beam data from the E852 and VES experiments on the $\pi\pi\pi$ final state showed important sensitivity to the number of partial waves allowed in fits to statistically limited datasets, with the strength of a $1^{-+}$ contribution through the $\pi \rho$ isobar being somewhat unclear. With the modern high-statistics dataset in COMPASS it has been possible to perform systematic studies of the waveset used in partial wave analysis, leading to a more reliable conclusion in which an enhancement compatible with $\pi_1(1600)$ is present, which is claimed to be definitively resonant~\cite{COMPASS:2021ogp}\footnote{At least when the momentum transferred to the proton in the process $\pi p \to (\pi\pi\pi) p$ is large, at smaller momentum transfer it is argued that non-resonant effects dominate.}.
Analysis of higher multiplicity final states in COMPASS is ongoing -- preliminary results presented in conferences show peaks in $b_1 \pi,\,  \rho \omega \to \pi \pi \omega$ and $f_1 \pi,\,  \rho a_0 \to \pi \pi \pi \eta$ that may be due to the $\pi_1(1600)$. 

\smallskip
Outside of production in hadron beam experiments, indications for a $\pi_1$ resonance were seen in the $\pi \eta'$ mode in the charmonium decay ${\chi_{c1} \to \eta' \pi \pi}$ at CLEO-c~\cite{CLEO:2011upl}. Preliminary analysis of this same decay at BESIII with large statistics, presented at the Hadron 2025 conference, has a strong $\pi_1(1600)$ contribution. $\chi_{c1} \to \eta \pi \pi$ was measured at CLEO-c, and more recently with much higher statistics at BESIII~\cite{BESIII:2016tqo}, where there is no strong evidence for a $1^{-+}$ $\pi \eta$ wave.

\bigskip
\bigskip

Theoretical expectations for the hadronic decay properties of hybrid mesons were developed within extensions of the `QCD-inspired' models described earlier, typically by assuming some quark pair-production dynamics applied in first-order perturbation theory (see the review article, Ref.~\cite{Meyer:2015eta}, for more details). One item of `lore' which originated in these studies was that hybrid mesons would have dominant quasi-two-body decays to meson-meson final states in which one meson has its $q\bar{q}$ pair in an $S$--wave, while the other is in a $P$--wave. This would suggest that the $\pi_1$, for example, should have much higher probability to decay into $\pi b_1$ than $\pi \rho$, despite the latter having much larger phase-space. That the supposedly preferred decay mode $\pi b_1$ ultimately leads to a complicated $5\pi$ final state made this hypothesis hard to test experimentally.

\smallskip
More recently, developments in lattice QCD have been such that it has been possible to consider excited hadrons truly as short-lived \emph{resonances} decaying to hadron-hadron pairs, and after being established as reliable for low-lying conventional mesons, the technology has been applied to the exotic $1^{-+}$ channel.
The approach, which relates the discrete spectrum of energy eigenstates in the finite volume of the lattice to hadron-hadron scattering amplitudes, is described in Chapter 20019 ``Lattice QCD calculations of hadron spectroscopy'' [\url{https://arxiv.org/abs/2505.10002}]. In principle it should be faithful to the full QCD physics, giving rise to coupled-channel scattering amplitudes in which resonances appear as poles at a complex value of the scattering energy~\cite{Briceno:2017max}.

\smallskip 

Actual calculations have seen this applied, typically with unphysically heavy quark masses, to \emph{elastic} resonances (such as the $\rho$ resonance in single-channel $\pi\pi$ scattering) 
and \emph{coupled-channel} resonances. An example of the latter case finds two light isoscalar tensor mesons, \mbox{$f_2$, $f_2'$} with $\pi \pi$, $K\bar{K}$ decays~\cite{Briceno:2017qmb}, broadly compatible with the experimental $f_2(1270), f_2(1525)$ states. An interesting feature of this calculation which has relevance for the empirical `lore' used in searches for glueballs is that we see something resembling the OZI--rule emerging from a first-principles QCD calculation.

\smallskip
This lattice finite--volume technology is currently limited to considering any number of coupled hadron-hadron scattering channels, but not to situations where channels of any higher multiplicity are kinematically accessible\footnote{But progress is being made in extending the formalism, see Ref.~\cite{Hansen:2019nir}.}.
The experimental $\pi_1(1600)$ lies above a number of many-hadron thresholds, so that with current formalism there is no practical way to investigate it as a resonance in a lattice QCD calculation using physical quark masses. In order to learn something about a $1^{-+}$ state in QCD, with a reasonable level of rigor, a first lattice calculation was performed considering QCD with three flavors of quark all set to the physical strange quark mass~\cite{Woss:2020ayi}. This choice leads to two dramatic simplifications: the explicit exact SU(3) flavor symmetry reduces the number of independent scattering channels, while the heaviness of the pion and other light hadrons ensures that the $1^{-+}$ state lies above only meson-meson thresholds, and below any three-meson thresholds.

\smallskip
The calculation in Ref.~\cite{Woss:2020ayi} computed spectra across six lattice volumes, constraining a coupled-channel system featuring channels that correspond (in the SU(3) flavor limit) to $\pi \eta, \pi \eta', \pi \rho, \pi b_1, \pi f_1, \rho \omega$, leading to amplitudes found to have a single resonance lying well above the $\pi \eta, \pi \rho$ thresholds, and close to the $\pi b_1, \pi f_1$ thresholds. The resonance was found to be narrow, with small couplings to all kinematically open channels, and a large coupling to the kinematically closed $\pi b_1$ channel. This observation is conceptually compatible with the older `lore' developed from models.

\smallskip

\begin{figure}[h]
	\centering
	\includegraphics[width=.7\textwidth]{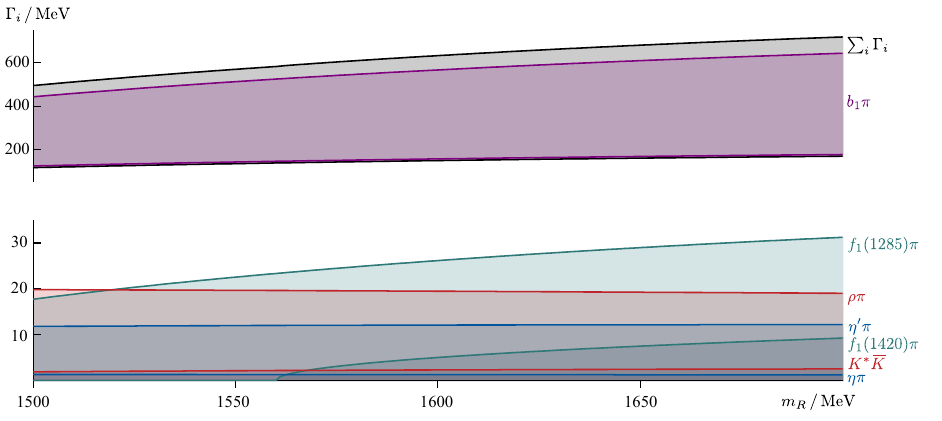}
	\caption{Partial widths into various quasi-two-body decays modes as a function of the $\pi_1$ resonance mass. Determined using extrapolation of the SU(3) flavor point lattice QCD results of Ref.~\cite{Woss:2020ayi}. The width of the colored bands indicate the degree of uncertainty due to the computed couplings. The total width, obtained by summing the partial widths, is shown by the grey band. Figure taken from Ref.~\cite{Woss:2020ayi}.}
	\label{fig:LQCDpi1}
\end{figure}

Previous lattice QCD calculations at various unphysical values of  quark mass indicate a plausible extrapolation scheme where `scaled' dynamical couplings remain constant with varying quark mass, but where kinematic quantities are replaced with the physical values\footnote{The couplings are assumed to scale like $c^\mathrm{phys} = \left| \tfrac{k^\mathrm{phys}(m_R^\mathrm{phys})}{k(m_R)} \right|^\ell \cdot c$ for a decay of a resonance of mass $m_R$ with decay momentum $k$ in partial-wave $\ell$. The phase-space for each decay must also be adjusted to correspond to physical kinematics.}. Performing this extrapolation leads to the partial decay widths shown in Figure~\ref{fig:LQCDpi1} as a function of the (assumed) mass of the $\pi_1$. If this extrapolation is reliable it indicates a broad resonance having the vast majority of its decays into $\pi b_1$. While somewhat imprecise, the large total width would appear to be in agreement with the $\pi_1$ determined in the analyses of Refs.~\cite{JPAC:2018zyd, Kopf:2020yoa}, but whether the dominance of $\pi b_1$ is accurate is yet to be seen experimentally.

\bigskip

Apart from the $\pi_1(1600)$, there is currently very little experimental evidence for further exotic hybrid meson candidates. One recently observed possibility is the $\eta_1(1855)$ claimed to be present in the charmonium radiative decay $J/\psi \to \gamma \, \eta \eta'$ at BESIII~\cite{BESIII:2022riz}. Analysis of the data suggests a relatively narrow isoscalar resonance decaying to $\eta \eta'$ with $J^{PC} = 1^{-+}$. In principle this could represent the first evidence for one of the flavor partners of the $\pi_1(1600)$ -- recall Figure~\ref{fig:LQCDmesons} in which two isoscalar states appear above the lightest isovector $1^{-+}$, with the lighter of the two being dominated by hidden light-quark structure and the heavier having dominantly hidden-strange structure.
In fact the observation of only a single $1^{-+}$ resonance in $J/\psi \to \gamma \, \eta \eta'$ poses something of a mystery. The production process is assumed to proceed through an intermediate gluonic state that must be in an SU(3) flavor singlet, suggesting that the produced resonance has at least some singlet component. On the other hand the $\eta$ meson is close to being a flavor octet, and the $\eta'$ close to being a flavor singlet, so that the $\eta \eta'$ system is close to flavor octet, and the resonance must have at least some octet component. Hence the $\eta_1(1855)$ must be an admixture in this flavor space (as are both of the isoscalar states in Figure~\ref{fig:LQCDmesons} discussed above). It then is a surprise that the orthogonal admixture, a second $\eta_1$ state, either heavier or lighter, is not observed in the analysis of the BESIII data. We await experimental confirmation of the $\eta_1(1855)$ in other processes.

\smallskip

There are no experimental candidates for positive parity exotic $J^{PC}$ resonances, and neither are there (to date) any first-principles QCD predictions for their likely decay modes. 
Searches for hybrid mesons with non-exotic $J^{PC}$ in principle share many of the problems discussed for glueballs. There can be mixing of basis states between hybrid and conventional $q\bar{q}$ states to form the physical eigenstates (although the large overlaps onto the chromomagnetic operators in the lattice study discussed above appear to indicate that dominant hybrid content can be assigned to certain states) and we lack any obvious distinguishing feature measurable in experiment for the hybrid nature. This takes us back to state-counting arguments as for glueballs, but examination of the literature leads us to rapidly conclude that excited states in the $0^{-+}$ and $1^{--}$ channels are quite poorly constrained experimentally in general.

\smallskip

A new experimental tool that promises to offer novel insights into the light meson spectrum is the GlueX spectrometer at Jefferson Lab, that is collecting a huge data set of meson photoproduction (see Chapter 60014 ``The GlueX Experiment''). They have recently used a subset of their data to estimate an upper limit of possible $\pi_1$ production~\cite{GlueX:2024erj} in advance of full partial wave analyses.

\section{Hybrid Baryons}\label{sec:hybridbaryons}

Hybrid baryons have attracted less attention than their meson cousins. In large part this is due to there being no exotic quantum numbers in the baryon sector, where all half-integral $J^P$ can be reached with quark model $qqq$ structures. As such there is no smoking gun signal, and while one could look for supernumerary states, there is also the longstanding `missing resonances' problem, where the set of experimentally observed baryon resonances is significantly smaller than the set expected within quark models~\cite{Crede:2013kia}.

\begin{figure}[h]
	\centering
	\includegraphics[width=.95\textwidth]{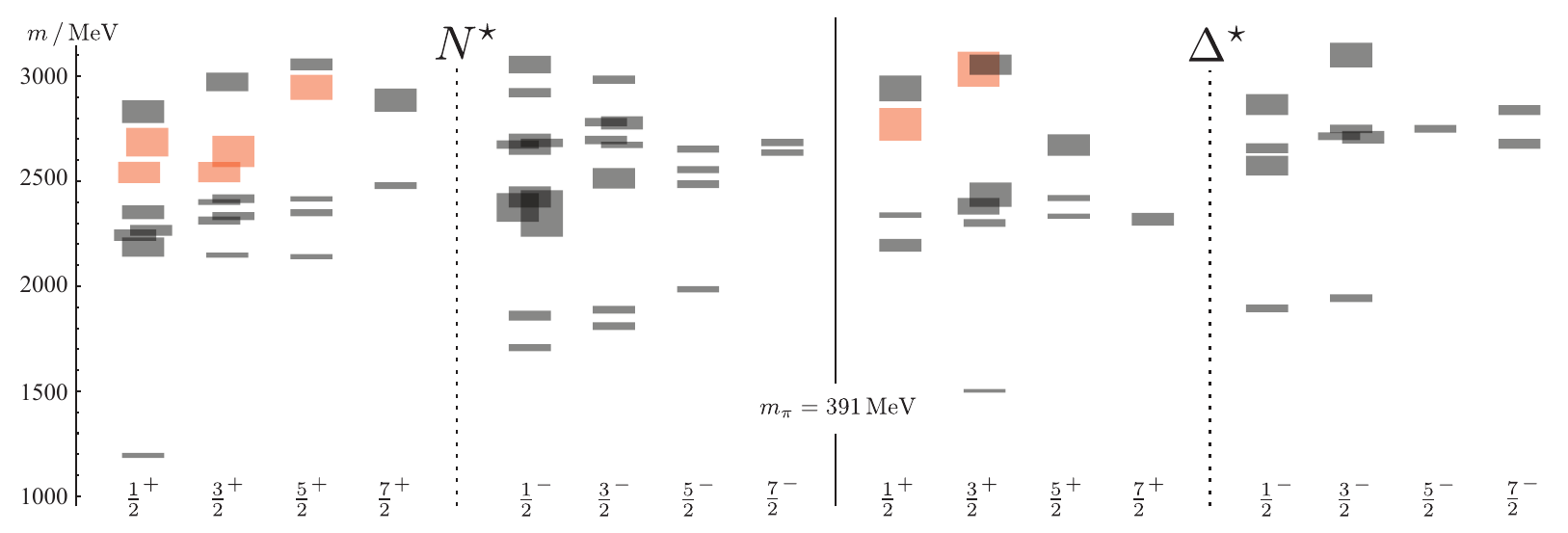}
	\caption{Lattice QCD computed spectrum of nucleons and Delta baryons with light quarks such that $m_\pi = 391$ MeV. States colored orange are found to have strong overlap onto operators containing a chromomagnetic gluonic structure, and are considered to be candidates to be \emph{hybrid baryons}. Figure adapted from Ref.~\cite{Dudek:2012ag}.}
	\label{fig:baryons}
\end{figure}

\smallskip
As was the case for hybrid mesons, older models make disparate predictions for the spectrum of hybrid baryons, with common suggestions being that there would be additional nucleon ($I=\tfrac{1}{2}$) and $\Delta$ ($I=\tfrac{3}{2}$) states with low spins and positive parity.
Lattice QCD results appeared in 2012 using a technique analogous to that discussed above for mesons, making use of a large basis of operators built from three quark fields and gauge-covariant derivatives, respecting the antisymmetry required for fermion fields~\cite{Dudek:2012ag}. Figure~\ref{fig:baryons} presents the spectrum of nucleon and $\Delta$ states extracted on a lattice with light quark masses such that the pion mass was $391$ MeV.
The distribution of states (shown in gray) is broadly consistent with expectations of $qqq$ quark models containing low angular momentum excitations (more discussion of these `conventional' baryon excitations can be found in Ref.~\cite{Edwards:2011jj}) while other states (shown in orange) don't fit into this pattern. These states are found to have strong overlap onto operators featuring the same chromomagnetic operator (commutator of gauge-covariant derivatives) that appeared in the meson case, and as such serve as candidates to be hybrid baryons.

\begin{figure}[h]
	\centering
	\includegraphics[width=.75\textwidth]{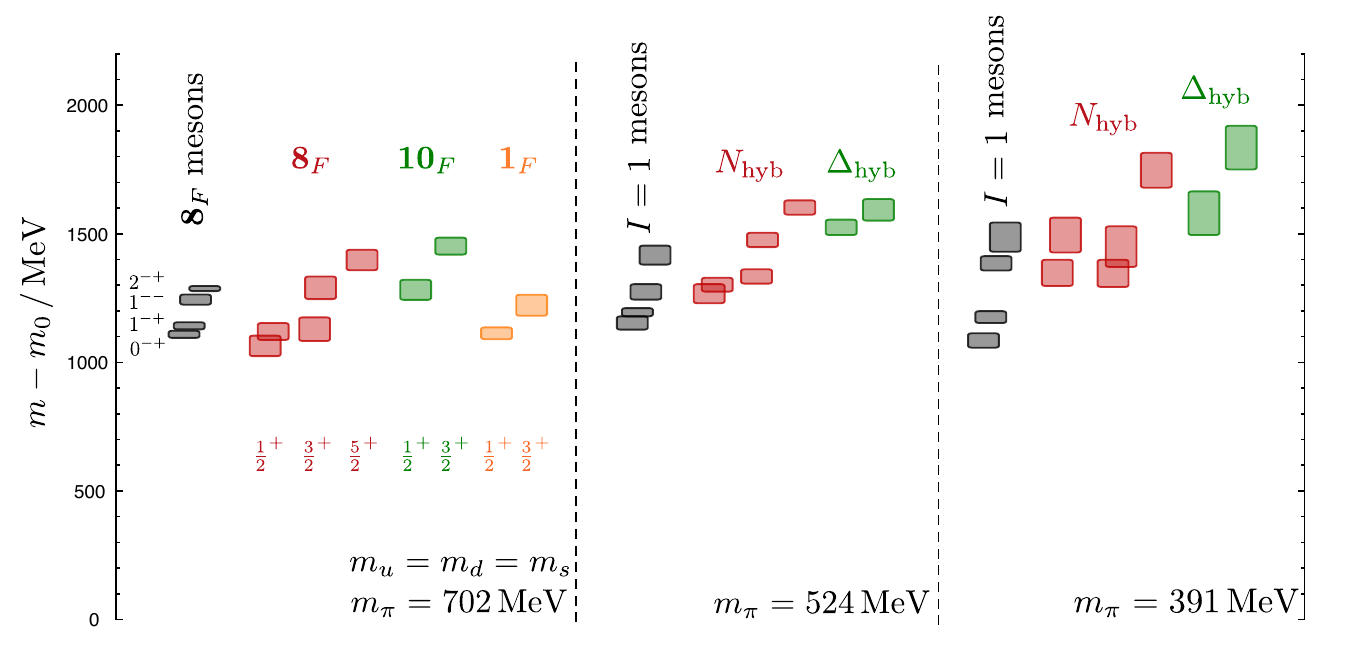}
	\caption{Spectrum of lightest hybrid mesons (with $\rho$ mass subtracted) and hybrid baryons (with nucleon mass subtracted) computed in lattice QCD. Three quark masses shown, left panel in SU(3) flavor limit with baryons labelled by their SU(3) flavor representation. The common energy scale of states may be interpreted as the energy associated with the chromomagnetic gluonic excitation. Figure adapted from Ref.~\cite{Dudek:2012ag}.}
	\label{fig:chromomagnetic}
\end{figure}

It was immediately noted that the distribution of states in $J^P$ for the lattice QCD hybrid baryons, together with the common operator overlap, suggested a relationship to the lattice QCD hybrid mesons. Having the hybrid baryons being three quarks, all in $S$-wave in a color octet $(qqq)_\mathbf{8}$, with the color neutralized by a $J^P = 1^+$ chromomagnetic excitation, explains the observations. The results also suggest that the nature of the gluonic excitation may be common to both mesons and baryons. As shown in Figure~\ref{fig:chromomagnetic}, an estimate of the energy associated with the gluonic excitation comes from subtracting a baseline energy due to the quarks, in the case of the mesons this is chosen to be the $\rho$ mass, and for the baryons, the nucleon mass.
While there is some `fine-structure' separation of states with spin, on average the excitation energy in baryons appears to be the same as in mesons, at roughly 1.2 GeV, and seems to be largely independent of the light quark mass used in the calculation.

\bigskip

Experimentally, comparing the PDG spectrum with Figure~\ref{fig:baryons}, in each of the relevant cases: $N_{1/2^+}, N_{3/2^+}, N_{5/2^+}$, and $\Delta_{1/2^+}, \Delta_{3/2^+}$, the number of established resonances does not yet even account for the number of expected conventional (gray) states lying below the first hybrid (orange) states. As  such any attempt to identify hybrid baryon components in the spectrum is currently impractical.

\smallskip

One advantage the baryon sector has over the meson sector is that use of stable proton targets in electron beam experiments allows access to \emph{resonance transition form-factors} in electroproduction, $\gamma^* N \to N^\star \to N \pi, N \pi \pi, K Y \ldots\,$. These form-factors, which should be sensitive to the internal structure of the resonances, are expressed in terms of a dependence upon the virtuality of the photon, and it has been proposed that hybrid baryons might have a different $Q^2$ dependence to conventional baryon excitations (see for example Ref.~\cite{Li:1991yba}). 
To date there are no solid expectations based upon first-principles QCD for how this might manifest, but in principle it can be addressed with future lattice QCD computations.

\newpage
\section{Conclusions}\label{sec:conclusions}

A role for gluonic excitations in the spectrum of hadrons would appear to be a natural consequence of strongly coupled non-Abelian QCD, but elucidating this role within non-perturbative QCD, and in the experimental spectrum, is an ongoing challenge.

\smallskip
Glueballs, states of pure glue requiring no quark involvement, have been shown numerically to feature in the quarkless Yang-Mills theory, but the lightest such states have quantum numbers also accessible to $q\bar{q}$ mesons. The likely admixture of glueballs with $q\bar{q}$ configurations has led to searches for supernumerary states, and the leading experimental candidate states include three scalar meson resonances lying between 1 and 2 GeV, although whether any one of these states is dominantly glueball in nature is undecided.
Hybrid mesons, in which an excitation of the gluonic field is coupled to $q\bar{q}$, can appear with \emph{exotic} $J^{PC}$, not accessible to a simple $q\bar{q}$ pair. Over time increasing levels of experimental evidence have been collected for a broad $1^{-+}$ isovector resonance, the $\pi_1(1600)$, which has properties that may be in agreement with lattice QCD calculations. 
The existence of hybrid mesons suggests that we might also have hybrid baryons, with plausibly the same gluonic excitation, but coupled to a $qqq$ system. There are no exotic half-integral $J^P$, and the current state of understanding of the experimental baryon spectrum makes searches for these states very challenging.

\smallskip
Identifying gluonic excitations within the experimental spectrum relies upon phenomenology based upon expectations coming from the structure of QCD, intuition based upon empirical observations, and features of models. Much `lore' has been developed, and commonly accepted, not all of it having a strong foundation in QCD

\smallskip
Lattice QCD appears to be the theoretical tool offering the most hope for first-principles understanding of QCD, but the calculations are extremely challenging. Since all the dynamics of QCD is present, calculations should yield a faithful representation of `all the physics' of the hadron spectrum, and this means that excited states will appear as \emph{resonances} in scattering just as they do in experiment. Of course we know that for physical values of the quark masses, many of the excited states we are interested in lie above many multi-hadron thresholds, and with current techniques all such channels must be considered in the analysis of lattice QCD finite-volume spectra. At this time, two-hadron decay channels can be handled straightforwardly, but three-hadron decays are at the cutting edge.
Initial progress will likely come from simplifying the system in a controlled manner: by using heavier than physical light quarks, multi-particle thresholds are moved up to higher energy and more excited states can be considered having only two-body decays; by setting the light quark masses equal to the strange quark mass, an exact SU(3) flavor symmetry simplifies the channel space still more~\cite{Dudek:2024roh, Johnson:2020ilc, Woss:2020ayi}.
Interrogating the extracted resonances to expose their internal structure requires further developments, but ideas include using currents external to QCD as probes.
The goal would be to establish a spectrum of excited states in QCD, make some proposals for their structure, and develop some more reliable 'lore' about their decays coming directly from strongly-coupled QCD, and build a phenomenology based upon this with which to view experimental observations.

\smallskip
Determining a definitive spectrum of resonances in each $J^{PC}$ is a necessary precursor to identifying the structure of each excited state, and this can be achieved by application of \emph{coupled-channel} analysis to as much experimental data as possible, obtained in a range of production processes.

\begin{ack}[Acknowledgments]%

 \smallskip
The author acknowledges support from the U.S. Department of Energy contract DE-SC0018416 at William \& Mary. This work contributes to the goals of the U.S. Department of Energy \emph{ExoHad} Topical Collaboration, Contract No. DE-SC0023598.
\end{ack}


\bibliographystyle{unsrt}
\bibliography{draft}

\end{document}